\documentclass[a4paper]{jpconf}
\usepackage{amsmath,amssymb,graphicx,bm}
\usepackage{color}
\usepackage[bottom]{footmisc}

\newcommand{\beq}{\begin{equation}}
\newcommand{\eeq}{\end{equation}}
\newcommand{\bea}{\begin{eqnarray}}
\newcommand{\eea}{\end{eqnarray}}
\newcommand{\ba}{\begin{align}}
\newcommand{\ea}{\end{align}}
\newcommand{\bfig}{\begin{figure}}
\newcommand{\efig}{\end{figure}}

\newcommand{\D}{\displaystyle}

\newcommand{\tplus}{t_{+}}

\begin{document}
\title{The $D\pi$ form factors from analyticity and unitarity}

\author{B.Ananthanarayan$^{a,\ast}$, I.Caprini$^{b,\dagger}$, I.Sentitemsu Imsong$^{a,\ddagger}$}

\address{$^a$ Centre for High Energy Physics,
Indian Institute of 
Science, Bangalore 560 012, India \\
$^b$ Horia Hulubei National Institute for Physics and Nuclear Engineering,
P.O.B. MG-6, 077125 Magurele, Romania}

\ead{$^{\ast}$anant@cts.iisc.ernet.in \\
\hspace{1.2cm}$^{\dagger}$caprini@theory.nipne.ro \\
\hspace{1.2cm}$^{\ddagger}$senti@cts.iisc.ernet.in}

\begin{abstract}
We study the shape parameters of the $D\pi$ scalar and vector form factors using as input
dispersion relations and unitarity for the moments of suitable heavy-light correlators evaluated 
with Operator Product Expansions, including $O(\alpha_s^2)$ terms in perturbative QCD.
For the scalar form factor, a low energy theorem and phase information on the unitarity cut are implemented  
to further constrain the shape parameters. We finally determine points on the real axis and isolate regions in the 
complex energy plane where zeros of the form factors are excluded.
\end{abstract}

\section{Introduction} \label{sec:intro}
\bigskip

The knowledge of shape parameters of the $D\pi$ form factors is of interest for 
the determination of the element $|V_{cd}|$ of the Cabibbo-Kobayashi-Maskawa (CKM) 
matrix entering precision tests of the Standard Model. The branching 
fractions of the semileptonic decays  $D\rightarrow \pi l \nu$ and 
$D\rightarrow K l \nu$ have recently been analysed by the CLEO collaboration  \cite{Ge:2008yi,Besson:2009uv}, 
which renewed the interest in the theoretical study of these processes.

The knowledge on the form factors can be further improved using analyticity 
and unitarity techniques, in particular, the method of unitarity bounds \cite{Okubo,SiRa}.  
Using this framework, one can obtain bounds on the form factors using as
input an integral of the modulus squared of the form factor along the
unitarity cut, known from unitarity and a dispersion relation for a
suitable correlator of the same current, which can be evaluated by
Operator Product Expansion (OPE) in the spacelike region.
Using standard mathematical techniques, one can then 
correlate the values of the form factor and its derivatives at 
different points inside the analyticity domain. 
A review of the method was presented recently in \cite{Abbas:2010jc}.
In this work, we apply  the above technique to the $D\pi$ form factors. The work is
motivated in part by the progress in perturbative QCD calculations, 
which now yield the heavy-light correlators of interest to order $\alpha_s^2$  
\cite{Chetyrkin:2001je} for a massless light quark and partly form the new experimental analyses 
of the semileptonic $D$ decays.

We investigate the shape parameters entering the Taylor expansion around $t=0$,
\beq
       f_{k}(t) = f_k(0)\left(1 +\lambda_{k}'\frac{t}{M_\pi^2} + \frac{1}{2} \lambda_{k}''\frac{t^2}{M_\pi^4} + \cdots\right), ~k=+,0,
       \label{eq:taylor}
\eeq 
and derive allowed ranges for the slopes  $\lambda_{k}'$ and the curvatures $\lambda_{k}''$.
Here, $f_{+}(t)$ and $f_{0}(t)$ are respectively the vector and scalar form factors and are related to each other
by the relation,
\beq\label{eq:f0}
f_0(t)=f_+(t)+\frac{t}{M_D^2-M_\pi^2} f_-(t), \quad t=q^2=(p-p')^2.
\eeq
 
We further find points on the real $t$-axis and isolate regions in the complex 
$t$-plane where zeros of the form factors are excluded. 
The knowledge of the zeros are of interest, for instance, for the dispersive methods 
based on phase (Omn\`es-type representations) and for testing specific models of the form factors.


\section{Review of the method}\label{sec:method}

\bigskip
To start with, we have the heavy-light invariant amplitudes $\Pi_+(q^2)$ and $\Pi_0(q^2)$ 
defined by the vector-vector correlation function,
\begin{equation}\label{eq:corr}
-(q^2 g^{\mu\nu} - q^\mu q^\nu)\Pi_+(q^2) + q^\mu q^\nu \Pi_0(q^2)
= i\int d^4x e^{iqx} \langle 0|TV^\mu(x) V^{\nu\dag}(0)|0 \rangle, 
\end{equation}
where $V_\mu=\bar d\gamma_\mu c$.

We then consider the moments of the invariant amplitudes at $q^2=0$, 
\begin{equation}\label{eq:chin}
\chi_{k}^{(n)} \equiv \frac{1}{n!}\frac{d^{n} }{(dq^2)^n}\left[\Pi_{k}(q^2) \right]_{q^2=0}, \quad k=+,0, \end{equation}
which satisfy dispersion relations of the form,
\begin{equation}\label{eq:chindr}
\chi_{k}^{(n)} =\frac {1}{\pi} \int_{t_+}^\infty\! dt\,\frac{{\rm Im}\Pi_k (t+i\epsilon)}{t^{n+1} },  \quad k=+,0,
\end{equation}
where $t_\pm=(M_D \pm M_\pi)^2$.

It is well-known from QCD that the amplitude $\Pi_+(q^2)$ satisfies a once subtracted dispersion 
relation, while for  $\Pi_0(q^2)$ an unsubtracted relation converges.  
The quantities  $\chi_{+}^{(n)}$  and $\chi_{0}^{(n)}$ are hence defined for $n\ge 1$ and  
$n\ge 0$, respectively. These relations are connected to the form factors $f_+(t)$ and $f_0(t)$  
by means of unitarity, 
\beq\label{eq:unit+}
{\rm Im}\Pi_+(t+i\epsilon) \geq \frac{3}{ 2}\frac{1}{ 48 \pi} \D \frac {\left[(t-t_+)(t-t_-)
\right]^{3/2}}{t^3} |f_+(t)|^2,
\eeq
\beq\label{eq:unit0}
{\rm Im} \Pi_0(t+i\epsilon) \ge \frac{3}{2} \frac{t_+ t_-}{ 16\pi}
\frac{[(t-t_+)(t-t_-)]^{1/2}}{ t^3} |f_0(t)|^2,
\eeq
respectively for the vector and scalar form factor, which hold for $t>t_+$. The above expressions are the unitarity 
sums for the spectral functions, restricted to the contribution of the $D\pi$ states in the isospin limit. 

OPE on the other hand allows for the calculation of $\chi_{k}^{(n)}$
as the sum of  perturbative and nonperturbative contributions.
While the perturbative parts of the moments of heavy-light correlators for $n\leq 7$ were 
calculated  up to order $\alpha_s^2$  in \cite{Chetyrkin:2001je}, the leading non-perturbative
contribution of the quark and gluon condensates were obtained from \cite{Lellouch:1995yv},
the corresponding expressions of which are presented in \cite{Ananthanarayan:2011uc}. 
For the vector and scalar form factors, we specifically use the dispersion relations 
for the moments $\chi_+^{(1)},  \chi_+^{(2)}, \chi_+^{(3)}$ and $\chi_0^{(0)},  \chi_0^{(1)}, \chi_0^{(2)}$
respectively, as a result of which we obtain, for each form factor, a family of three different constraints. 
The final allowed domain for the parameters of interest will be the intersection of the three individual domains.

From the dispersion relations (\ref{eq:chindr}) and the unitarity conditions given in 
eqns.(\ref{eq:unit+}) and (\ref{eq:unit0}), it follows that each form factor $f_{k}(t)$ satisfies a set of 
integral inequalities written as,
\beq
 \frac{1}{\pi} \int^{\infty}_{\tplus } dt\ \rho_k^{(n)}(t) |f_{k}(t)|^{2} \leq \chi_k^{(n)},\quad \quad k=+,0,
        \label{eq:I}
\eeq
where the  weights $\rho_k^{(n)}(t)$ are the product of $1/t^{n+1}$ with the phase space factors entering 
the unitarity relations. Using the fact that the form factors $f_{+}(t)$ and $f_{0}(t)$ are analytic 
functions in the complex $t$-plane cut along the real axis from  $t_+$ to $\infty$, we apply
standard techniques to derive from eqn.(\ref{eq:I})  
constraints on their values, in particular on the shape parameters and on the regions  
in the real and complex energy plane where zeros are excluded. 
Specifically, we cast the problem  into a canonical form via a conformal
map, and derive a determinant which is central to our investigations for obtaining 
bounds on the shape-parameters \cite{Abbas:2010jc, Ananthanarayan:2011uc, Anant:2011}
and for finding the exclusion regions of the zeros. 
The quantity $f_k(0)$ defined in eqn.(\ref{eq:taylor}) goes as an important input to our work.
Besides, additional information for the scalar form factor provided by a 
low-energy soft-pion theorem namely the Callan-Treiman relation \cite{CallanTreiman},
is exploited to obtain more stringent bounds on the shape parameters of the scalar form factor
For the $D\pi$ case, the corresponding expression is given below,
\beq\label{eq:CT}
f_0(\Delta_{D\pi}) = f_D/f_\pi,
\eeq
where  $\Delta_{D\pi}= M_D^2-M_{\pi}^2$ is the relevant Callan-Treiman point  
and $f_D$ and $f_\pi$ are the meson decay constants.
 
Even more stringent constraints on the shape parameters can be obtained if we also have some 
information on $f_{k}(t)$ on the unitarity cut, in particular, if the phase  $\delta_k(t)$  defined as,
\beq\label{eq:phase}
f_k(t+i\epsilon)=|f_k(t)| {\rm e}^{i \delta_k(t)}, ~~k=+,0,
\eeq
is known along a low-energy interval, $t_+\leq t\leq t_{\rm in}$. This information on the phase 
can be implemented using the technique of generalized Lagrange multipliers and 
involves the  solution of an integral equation. A review of the method and more 
references are given in  \cite{Abbas:2010jc}.

\section{Inputs and Results \label{sec:input}}

\bigskip

The perturbative and nonperturbative contributions to the moments $\chi_k^n$
have been tabulated in \cite{Ananthanarayan:2011uc}. The value of the inputs used
for $f_k(0)$ and $f_0(\Delta_{D\pi})$ which reads $0.67_{-0.07}^{+0.10}$ and
$1.58 \pm 0.07$ respectively are obtained from a recent work based on analyticity \cite{Khodjamirian:2009ys}.
We roughly estimate the phase of the $D\pi$ form factors from 
the masses and  widths of  the resonances dominant at low energies, 
namely, from the relativistic Breit-Wigner parametrization given by, 
\beq	\label{eq:phasedelta}
\delta(t) =  \arctan \left(\frac{M_R\Gamma(t)}{M_R^2-t}\right),
\eeq
where $M_R$ is the mass  and $\Gamma(t)$ the energy-dependent width - 
\beq	\label{eq:gamma}
\Gamma(t) = \left(\frac{q(t)}{q(M_R^2)}\right)^{2 J+1} \frac{M_R}{\sqrt{t}}\,\Gamma_R,
\eeq
written in terms of the angular momentum  $J$, the width $\Gamma_R$ and the centre of mass momentum 
$q(t)= \sqrt{(t-t_-)(t-t_+)/4t}$.
 
The central values of the masses and widths of the lowest charged $D\pi$ vector and 
scalar resonances, namely the $D^*$ and $D_0^*$ respectively are listed in  \cite{PDG}.
We note that the vector resonance $D^*$ is very close to the threshold so that a reasonable 
expression of the phase cannot be obtained from a Breit-Wigner parametrization,
but in the scalar case we can assume that the 
phase $\delta_0(t)$ is reliably described by the expression (\ref{eq:phasedelta}) with $J=0$.  
For our analysis, we consider the phase up 
to the point $t_{\rm in}= (2.6 \,{\rm GeV})^2$,  close to the first inelastic $D\eta$ channel opening at 2.42 GeV.

\bfig[ht]	
	\begin{center}\vspace{0.cm}
	 \includegraphics[angle = 0, clip = true, width = 2.3 in]
{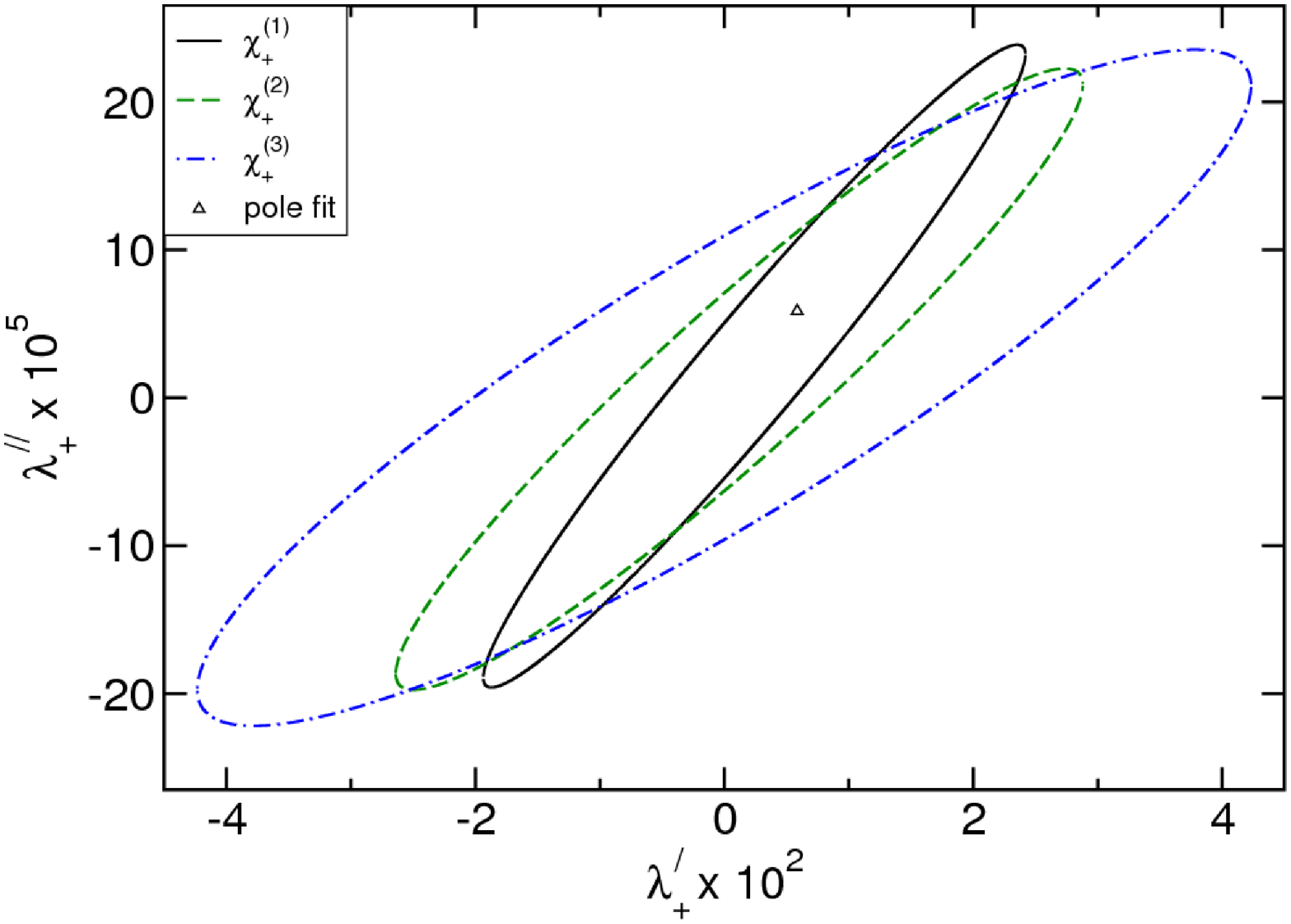}	
	\end{center}\vspace{-0.2cm}
	\caption{Constraints on the slope and curvature of the vector form factor obtained using as input 
different moments of the correlator. The point indicates the slope and curvature of the pole ansatz 
given in  \cite{Becirevic:1999kt}.}
	\label{fig:vec_slope_curv}
\efig

\bfig[ht]	
	\begin{center}\vspace{0.cm}
	 \includegraphics[angle = 0, clip = true, width = 2.3 in]
{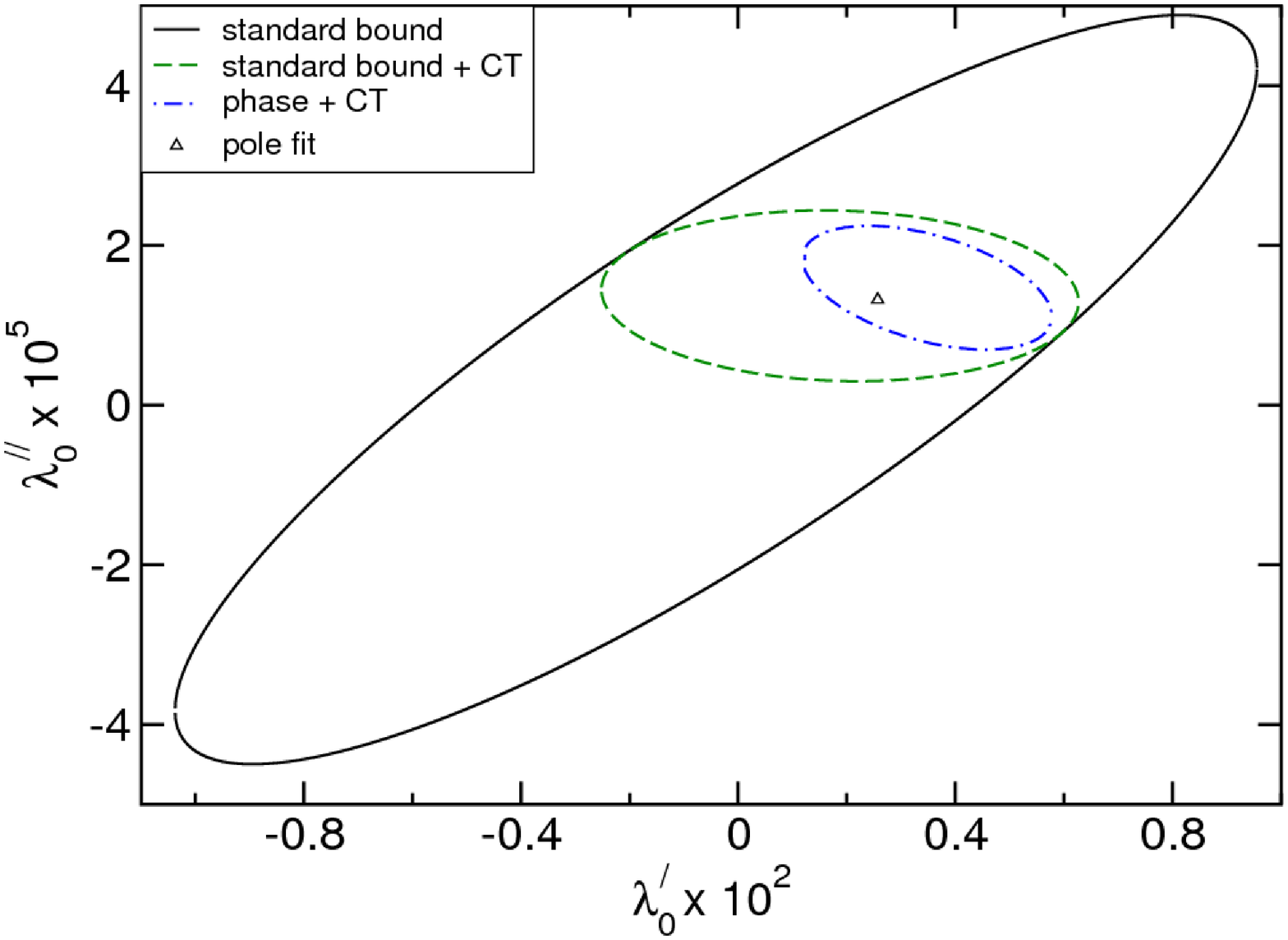}	
	\end{center}\vspace{-0.2cm}
	\caption{Constraints on the slope and curvature of the scalar form factor obtained with the 
moment $\chi_0^{(0)}$, from the standard bound and  by including the phase and the low-energy theorem. 
The point indicates the slope and curvature of the pole ansatz given in  \cite{Becirevic:1999kt}.}  
	\label{fig:scalar_slope_curv}
\efig	

The results of our analysis are presented in Fig.\ref{fig:vec_slope_curv}, where the
the interior of the ellipses represent the 
allowed domains in the slope-curvature plane for the vector form factor, obtained 
with  three moments $\chi^{(n)}_+$ of the vector correlator with just $f_+(0)$
and no other input. While the smallest or in other words the best domain is obtained from the lowest moment, 
we find that the higher moments tend to slightly reduce the domain, since one must take 
the intersection of all the domains in order to fulfill simultaneously the constraints.
We also indicate the slope and curvature of the simple pole ansatz \cite{Becirevic:1999kt},
which is seen to satisfy the unitarity constraints coming from our analysis.
In Fig.\ref{fig:scalar_slope_curv}, we show the bounds derived from 
the lowest moment $\chi_0^{(0)}$ of the scalar form factor
where we also use information on the phase and the soft pion theorem. 
The effect of these additional constraints can clearly be seen in the figure.
We also show the slope and curvature from the pole ansatz \cite{Becirevic:1999kt}
which satisfies our constraints. It may be noted that the above bounds were obtained  with the central 
values of the input parameters. By varying simultaneouly all the input values we can obtain 
more conservative regions, which are slightly larger than the domains shown in the figures.

\bfig[ht]	
	\begin{center}\vspace{0.3cm}
	 \includegraphics[angle = 0, clip = true, width = 2.3 in]
{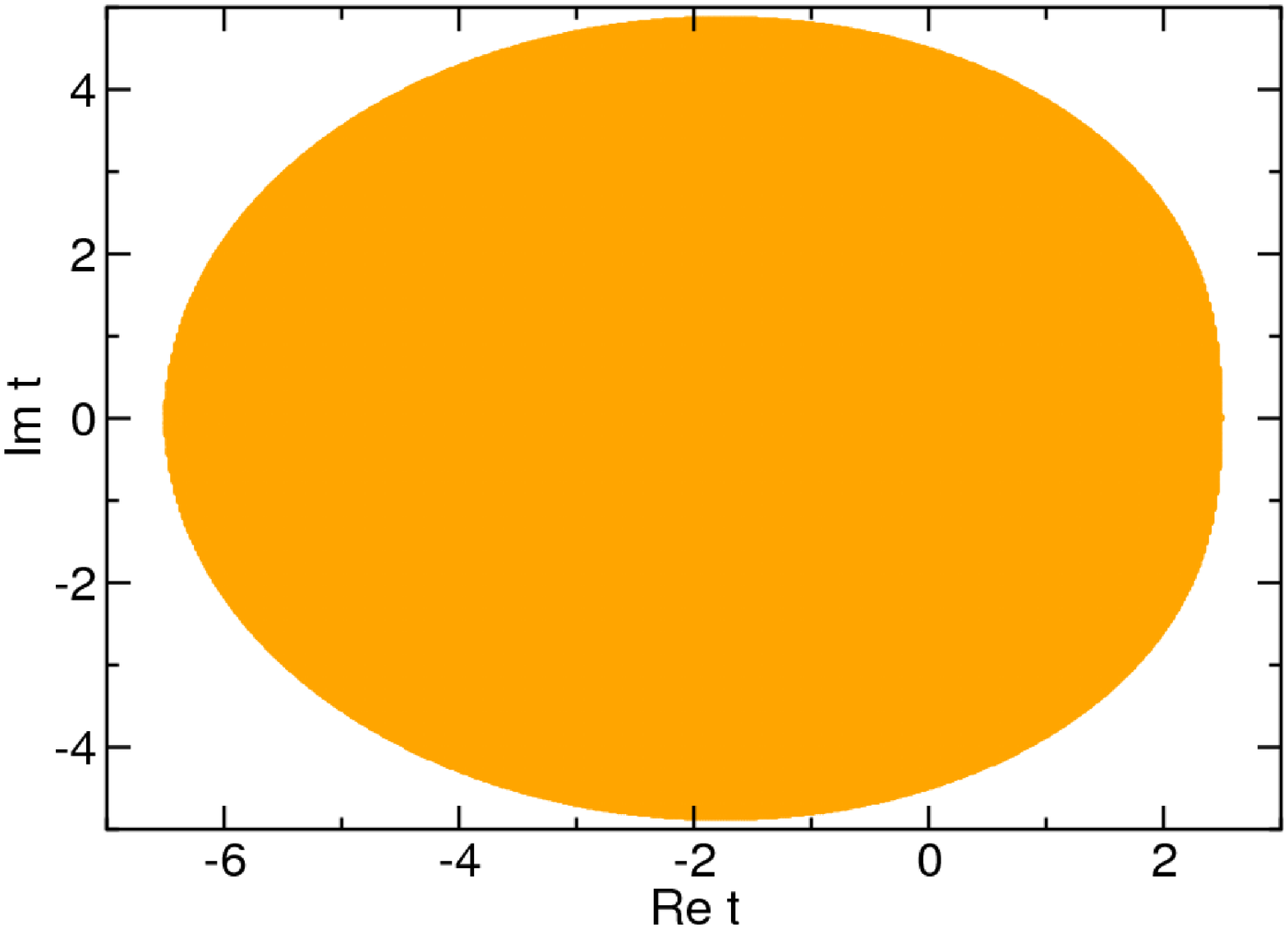}	
	\end{center}\vspace{-0.2cm}
	\caption{Domain without zeros for the vector form factor, obtained from the lowest moment 
$\chi_+^{(1)}$ and the input $f_+(0)$.}  
	\label{fig:vec_zeros}
\efig

We now present results on the zeros of the scalar and vector $D\pi$ form factors.
We have considered in the following only the constraints imposed by the 
lowest moments, corresponding to $n=1$ in the vector case and 
$n=0$ in the scalar one, using $f_+(0)$ as input quantity. 
For the vector form factor,real zeros are excluded in the range $(-1.0,\, 0.80)\,{\rm GeV}^2$. 
while for the scalar case it is $(-2.51,\, 1.55)\,{\rm GeV}^2$ for the standard bounds,  
With the inclusion of the low-energy constraint in the case of the scalar form factor,
the range is increased to  $(-3.55,\, 3.89)\,{\rm GeV}^2$. We present in Figs. \ref{fig:vec_zeros} and \ref{fig:scalar_zeros}   
the regions of excluded zeros in the complex $t$-plane for the vector and
scalar form factors respectively. In the scalar case the zeros are excluded in a larger domain if the 
low-energy constraint is also imposed. 

\bfig[ht]	
	\begin{center}\vspace{0.3cm}
	 \includegraphics[angle = 0, clip = true, width = 2.3 in]
{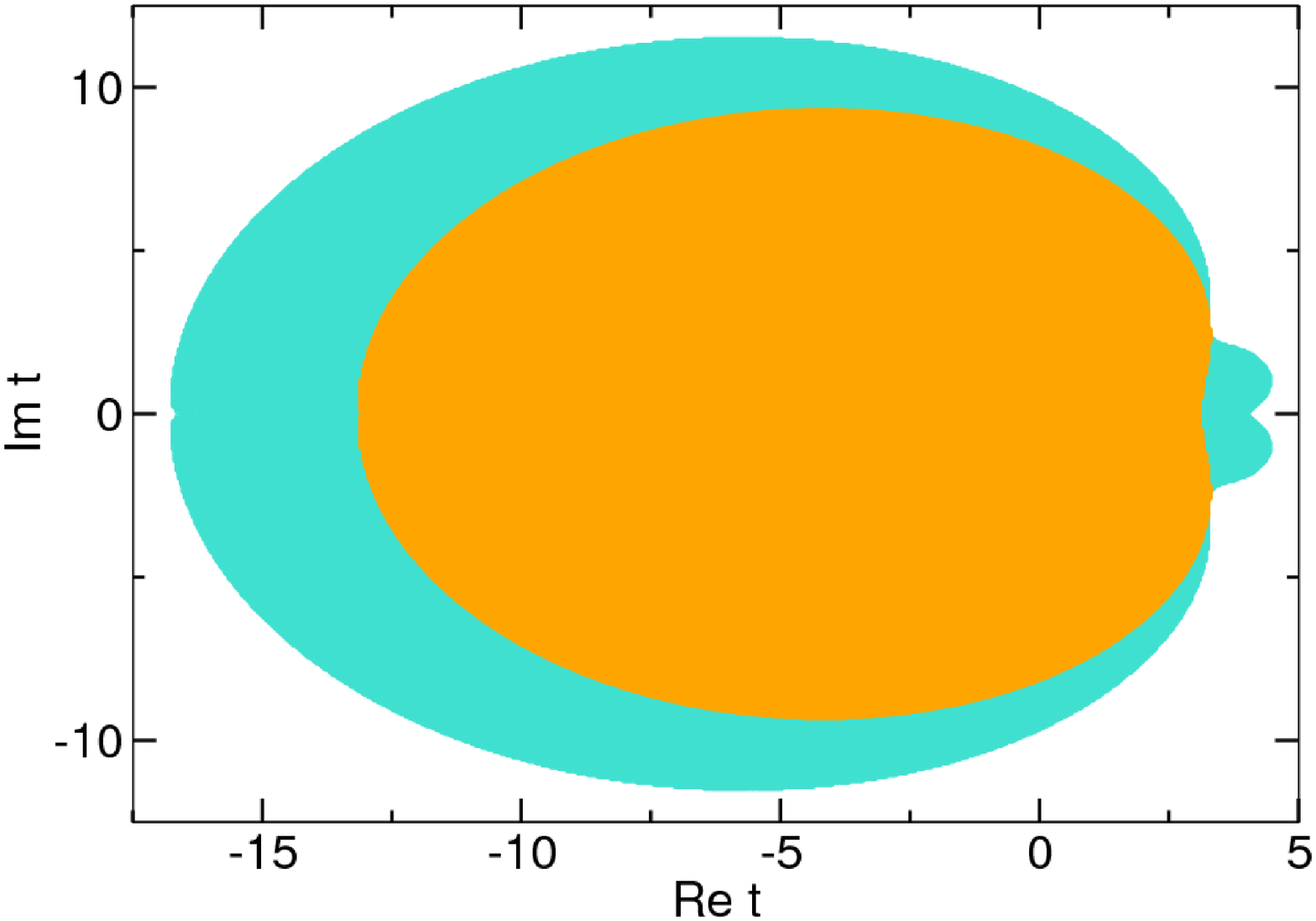}	
	\end{center}\vspace{-0.2cm}
	\caption{Domain without zeros for the scalar form factor, obtained from the lowest moment 
$\chi_0^{(0)}$ and the input $f_0(0)$ (smaller region) and using in addition the CT constraint (larger  region).}  
	\label{fig:scalar_zeros}
\efig

\section{Conclusions\label{sec:conc}}
\bigskip

In this work, we have used analyticity and unitarity techniques to derive 
information on the $D\pi$ form factors, which might be of interest for   
the determination of the element $|V_{cd}|$ of the CKM matrix.
We have applied the formalism of unitarity bounds and have derived a family of constraints 
to be satisfied by the shape parameters of both the scalar and vector form factors at $t=0$ 
and explored regions on the real axis as well as in the complex plane where the form factors 
cannot have zeros. The exclusion regions for the zeros that we have isolated basically 
cover a significant part of the entire low energy region. They are of practical interest, 
for instance, for the dispersive methods based on phase (Omn\`es-type representations) 
and for testing specific models of the form factors. 
The technique can be extended also for deriving new parameterizations
for the $D\pi$ form factors in the semileptonic region, similar to those
proposed in \cite{BoCaLe} for the semileptonic  $B \rightarrow \pi l\nu$ decay. The new
parametrizations properly implement the singularities related to the
lowest charmed resonances and are useful for the description of the
semileptonic data (see Ref.\cite{Ananthanarayan:2011uc} for more details).

\end{document}